\documentclass[12pt,aps,showpacs]{revtex4}
\usepackage{epsfig}
\usepackage{bbm}
\usepackage{amssymb}
\usepackage{amsmath}

\newcommand{\lbfig}[1]{\refstepcounter{fig} \label{#1} }
\newcounter{fig}
\newcommand{\nc}{\newcommand}
\nc{\be}{\begin{equation}}
\nc{\ee}{\end{equation}}
\nc{\bea}{\begin{eqnarray}}
\nc{\eea}{\end{eqnarray}}
\nc{\nn}{\nonumber}
\nc{\acom}[2]{ \left\{ #1,#2 \right\} }
\nc{\com}[2]{ \left[ #1,#2 \right] }
\nc{\dd}{^\dagger}
\nc{\ddp}{^{\dagger\prime}}
\nc{\ddpp}{^{\dagger\prime\prime}}
\nc{\pp}{^{\prime\prime}}
\nc{\ml}{M^\dagger M}
\nc{\mr}{MM\dd}
\nc{\explrp}{{\rm e}^{\frac{i}{2} \overset{\leftharpoonup}{\partial_z}
  \overset{\rightharpoonup}{\partial_{k_z}} }}
\nc{\exprlp}{{\rm e}^{\frac{i}{2} \overset{\leftharpoonup}{\partial_{k_z}}
  \overset{\rightharpoonup}{\partial_z} }}
\nc{\explrm}{ e^{-\frac{i}{2} \overset{\leftharpoonup}{\partial_z}
  \overset{\rightharpoonup}{\partial_{k_z}} }}
\nc{\exprlm}{ e^{-\frac{i}{2} \overset{\leftharpoonup}{\partial_{k_z}}
  \overset{\rightharpoonup}{\partial_z} }}
\nc{\lp}{\left(}
\nc{\rp}{\right)}
\nc{\CP}{{\cal CP}}
\nc{\CPd}{({\cal CP})^\dagger}
\nc{\Q}{{\cal Q}}
\nc{\Qd}{({\cal Q})^\dagger}
\nc{\col}{{\cal C}}
\nc{\cm}{{{\cal M}^2}}
\nc{\cs}{{\cal S}}

\def\Slash#1{#1\kern-0.55em\raise.05ex\hbox{/}}
\def\slash#1{#1\kern-0.5em\raise.05ex\hbox{{$\scriptstyle /$}}}

\begin{document}
\rightline{HD-THEP-05-04, ITP-UU-05/10, SPIN-UU-05/08}

%\vskip 0.2in

\title{
 MSSM Electroweak Baryogenesis
  and Flavour Mixing in Transport Equations}

\author{Thomas Konstandin$^*$, Tomislav Prokopec$^{(1)*}$,
    Michael G. Schmidt, Marcos Seco
       }
\email[]{T.Konstandin@thphys.uni-heidelberg.de}
\email[]{T.Prokopec@phys.uu.nl}
     %T.Prokopec@thphys.uni-heidelberg.de
\email[]{M.G.Schmidt@thphys.uni-heidelberg.de}
\email[]{M.Seco@thphys.uni-heidelberg.de}

\affiliation{Institut f\"ur Theoretische Physik, Heidelberg
University,
             Philosophenweg 16, D-69120 Heidelberg, Germany
}

\affiliation{$(1)\,$
             Institute for Theoretical Physics (ITF) \& Spinoza Institute,
             Utrecht University, Leuvenlaan 4, Postbus 80.195,
             3508 TD Utrecht, The Netherlands
             }

\date{\today}
\begin{abstract}

 We make use of the formalism developed in Ref.~\cite{fermions}, 
and calculate the chargino mediated baryogenesis in 
the Minimal Supersymmetric Standard Model. The formalism makes use of  
a gradient expansion of the Kadanoff-Baym equations for mixing fermions.
For illustrative purposes, we first discuss the semiclassical transport
equations for mixing bosons in a space-time dependent Higgs background. 
To calculate the baryon asymmetry, we solve a standard set of diffusion 
equations, according to which the chargino asymmetry is transported to the 
top sector, where it biases sphaleron transitions. 
At the end we make a qualitative and quantitative comparison of our results
with the existing work. 
We find that the production of the baryon asymmetry
of the Universe by CP-violating currents in the chargino sector
is strongly constrained by measurements of electric dipole moments.

\end{abstract}


\maketitle

%
%%%%%%%%%%%%%%%%%%%%%%%%%%%%%%%%%%%%%%%%%%%%%%%%%%%%%%%%%%%%%%%%%%%%%%%%%%%%%%%
%  MAIN TEXT
%%%%%%%%%%%%%%%%%%%%%%%%%%%%%%%%%%%%%%%%%%%%%%%%%%%%%%%%%%%%%%%%%%%%%%%%%%%%%%%
%

\section{Introduction}

Electroweak baryogenesis \cite{EWBG} is an effective framework for explaining
the baryon asymmetry of the universe (BAU). 
The most appealing feature of this mechanism lies in the fact
that the relevant physics will soon be explored by experiments,
most notably by LHC at CERN and by the new generation of
electric dipole measurements. 

It has been realized that the scenario of electroweak baryogenesis depends on
extensions of the Standard Model (SM), since two mandatory
conditions are not met in the SM.
The first reason is that CP-violation in the SM is marginal, such that
the observed magnitude of baryon asymmetry cannot be explained.
Secondly, the electroweak phase transition in the SM is a 
crossover~\cite{KajantieLaineRummukainenShaposhnikov:1996,
                CsikorFodorHeitger:1998},
leading to a too weak departure from equilibrium to be viable for
baryogenesis.

The Minimal Supersymmetric Standard Model (MSSM) instead
has all the necessary ingredients. CP violation is enhanced by adding
phases to the parameters in the soft supersymmetry breaking sector, which 
contribute to the chargino mass matrix. 
Furthermore, the additional bosonic degrees
of freedom can lead to a strong first order phase transition
as {\it e.g.} in the light stop 
scenario~\cite{CarenaQuirosWagner:1996,Bodeker:1996pc}.

 These considerations indicate that the MSSM has the potential
of explaining the observed BAU {\it via} electroweak baryogenesis.
However, a formalism that determines the baryon asymmetry has to incorporate
several features. Clearly the formalism has to reflect the quantum nature of
the involved particles, for CP violation is a purely quantum effect.
In addition, since the sphaleron processes are only operative in
the unbroken phase, the CP-violating particle densities have to be transported
away from the wall into the unbroken phase to lead to a net baryon density.
A formalism that can handle both of these aspects is given by the Kadanoff-Baym
equations, which are in turn derived from the 
out-of-equilibrium Schwinger-Dyson equations.

Early approaches that aimed to determine CP-violating densities
and have not attempted to derive transport equations from first principles
have been based on the dispersion relation of the quasi-particles
\cite{JoyceProkopecTurok:1995, ClineJoyceKainulainen:1998, Rius:1999zc,
ClineJoyceKainulainen:2000,HuberSchmidt:2000+2001}
deduced with the WKB method. For a recent resurrection of the method
see~\cite{BhattRangarajan:2004}. 

In \cite{Madrid_group, Madrid_group_imp} it was suggested that
an important contribution is given by mixing effects
of the quasi particles in the wall rather than from the dispersion 
relations in the case of a nearly degenerate mass matrix.
However, in the work \cite{Madrid_group, Madrid_group_imp} 
transport equations are not
derived in a first principle approach either, but
the current continuity equation is used to determine CP-violating
contributions to the Green functions in a perturbative
approach, which are subsequently inserted as sources into classical diffusion
equations derived in \cite{Huet_Nelson}. These classical diffusion
equations neglect oscillations of the off-diagonal elements of the
Green function that are important for a proper treatment of CP violation.

 Starting from the Kadanoff-Baym equations, 
the authors of~\cite{KPSW,dickwerk} have derived
the CP-violating semiclassical force in kinetic transport equations,
which appears in fermionic kinetic equation at second order in derivatives.
Initially, this was done for the one fermion flavour case~\cite{KPSW}
and then subsequently generalized to the diagonal part of the
multiflavour case~\cite{dickwerk}. 

Recently this formalism was advanced to include mixing fermions~\cite{fermions}.
The formalism provides an accurate description of the dynamics in
the thick wall regime, which applies to particles, whose de Broglie wave
length is much shorter than the thickness of the phase boundary (bubble wall),
formally $\partial_x \ll k$. 

One conclusion of the work \cite{fermions}
is that two features of the transport equations are not captured by
the procedure used in \cite{Madrid_group, Madrid_group_imp}. 
Firstly, the densities that are off-diagonal 
in the mass eigenbasis of the system will perform
oscillations analogously to neutrino oscillations. This effect 
suppresses the transport of the CP-violating sources, especially
if the mass spectrum
in the chargino sector is far from degeneracy. 
Secondly, while Refs.~\cite{Madrid_group, Madrid_group_imp} used a 
phenomenological prescription (Fick's law) to introduce the CP-violating
sources into the diffusion transport equations, 
no such prescription is required in our formalism. The sources enter
the diffusion transport equations with an unambiguously defined amplitude.

A first goal of this publication is to study the simpler bosonic case,
and thus to rectify the conclusions of~\cite{fermions}.
As a second  and principal goal, we consider the chargino mediated baryogenesis
in the MSSM, in order to study the effects of flavour oscillations
and source amplitude ambiguity on the baryon asymmetry 
within the framework of the reduced set of diffusion equations
for charginos and quarks used 
in~\cite{Madrid_group, Madrid_group_imp,BalazsCarenaMenonMorrisseyWagner:2004,Huet_Nelson}.
We also make a comparison of baryogenesis from the semiclassical
force mechanism. The principal difference with respect to the previous work,
is that our treatment is basis independent, while the calculations
presented in Ref.~\cite{ClineJoyceKainulainen:2000,KPSW,dickwerk} 
were performed in the mass eigenbasis. 

This article is organized as follows.
In section \ref{sec_bos} we derive transport equations for mixing bosons.
This is done mainly to clarify the conclusions from \cite{fermions} that
are present in the bosonic case, too.
In the subsequent section we discuss, how
the introduction of phenomenological damping terms can lead to additional
unphysical CP-violating sources. The sections \ref{sec_ferm} and
\ref{sec_diff} state the fermionic transport equations derived in
\cite{fermions} and the system of diffusion equations that is used to determine
the baryon asymmetry.
Numerical results are presented in section \ref{sec_num},
and we conclude in section \ref{sec_conc}.

\section{Transport Equations for Mixing Bosons \label{sec_bos}}

  In this section we will derive transport equations for mixing bosons
from the Kadanoff-Baym equations and the resulting CP violating
particle densities.
This is a simpler analogon to the derivation for
the fermionic case given in \cite{fermions}. In the fermionic case, the spinor
structure complicates the decoupling of the system of equations, but
 the bosonic case given here will already support the main
conclusions given in \cite{fermions} without the technical issues coming
from the spinor structure.

\subsection{Kadanoff-Baym Equations and the Approximation Scheme}

Starting point are the coupled Kadanoff Baym equations \cite{dickwerk}
\bea
&e^{-i\Diamond}\{k^2 - \cm \}\{\Delta^{<,>}\} - e^{-i\Diamond} \{\Pi_h\}\{\Delta^{<,>}\}
- e^{-i\Diamond} \{\Pi^{<,>}\}\{\Delta_h\} = \col, & \\
&\col = \frac{1}{2} \, e^{-i\Diamond} \lp \{\Pi^>\}\{\Delta^<\} - \{\Pi^<\}\{\Delta^>\} \rp,&
\eea
where $\Delta$ denotes the Green function and $\Pi$ the self-energy of the bosons.
Both quantities are $N\times N$ matrices in flavour space and depend on the
average coordinate $X_\mu$ and the momentum variable $k_\mu$.
The superscripts
$<,>$ and the subscript $h$ denote the additional
$2\times 2$ matrix structure as usual in the Kadanoff-Baym formalism

\be
\Delta =
\begin{pmatrix}
\Delta^{++} & \Delta^{+-} \\
\Delta^{-+} & \Delta^{--} \\
\end{pmatrix}, \nn
\ee
\bea
&\Delta^< = \Delta^{+-}, \quad \Delta^> = \Delta^{-+}, \quad
\Delta^t = \Delta^{++}, \quad \Delta^{\bar t} = \Delta^{--}, \quad &\nn \\
&\Delta_h = \Delta^t -\frac{1}{2} (\Delta^< + \Delta^>). &
\eea
The diamond operator coming from the transformation into Wigner space
is defined by
\be
\Diamond \{a\}\{b\} \equiv
        \frac{1}{2} \Big( (\partial_{X_\mu}a) \partial_{k_\mu} b
    - (\partial_{k_\mu}a) \partial_{X_\mu} b \Big).
\ee
The mass squared matrix $\cm$ is space-time dependent and hermitian.
During the electroweak phase transition, the bosonic particles relevant
for baryogenesis are the squarks whose mass matrix is given by
\be
{\cal M}^2 =
\begin{pmatrix}
m_Q^2 + h_t^2 H_2^2(X_\mu) & h_t(A_t H_2(X_\mu) - \mu_c^* H_2(X_\mu)) \\
h_t(A_t^* H_2(X_\mu) - \mu_c H_2(X_\mu)) & m_U^2 + h_t^2 H_2^2(X_\mu) \\
\end{pmatrix}.
\ee

In thermal equilibrium the Green function for a quasiparticle with mass $m$
is
\bea
i\Delta_{\rm eq}^<(k_\mu)
   &=& 2\pi \, \delta(k^2-m^2) \textrm{ sign}(k_0) f_{BE}(k_0) \nn \\
i\Delta_{\rm eq}^>(k_\mu)
  &=& 2\pi \, \delta(k^2-m^2) \textrm{ sign}(k_0)(1+ f_{BE}(k_0))
\label{delta_eq}
\eea
with the Bose-Einstein distribution function
\be
f_{BE}(k_0) = \frac{1}{e^{\beta k_0} - 1}.
\ee
The particle density can be deduced from the Green function using
\be
j_\nu(X_\mu) = 2i \int_{k_0>0} \frac{d^4k}{(2\pi)^4} \,  k_\nu \, \Delta^<(X_\mu, k_\mu).
\ee

Since there will be already a contribution to the CP-violating
particle densities in the mass term, we will in our approximation neglect
interactions with other particle species. However, we will keep the collision term $\col$,
since this term usually drives the system back to equilibrium and
allows to fulfill the physical boundary conditions far away from the wall. We
will not explicitly calculate the collision term, but finally replace it
by a phenomenological damping term. Hence the Kadanoff-Baym equations simplify to
\be
e^{-i\Diamond}\{k^2 - \cm \}\{\Delta^{<,>}\} = \col.
\ee
A further simplification is to perform the calculation in the bubble
wall frame. Our picture of the phase transition is as follows. Bubbles of
the Higgs field condensate nucleate and grow at a first order
electroweak transition, and as they become large,
they become approximately planar. The wall frame is then defined
as the frame moving with the bubble phase interface.
Due to the planarity, in this frame the mass matrix depends only on the
average coordinate $z:=X_3$.
In addition as mentioned in the introduction, we are working in the thick wall
regime, what makes a gradient expansion reasonable.
The system expanded up to first order in
gradients reads (prime denotes derivatives with respect to $z$):
\be
\lp k^2 + i k_z \partial_z + \frac{1}{4} \partial_z^2
- \cm - \frac{i}{2} \cm^\prime \partial_{k_z} \rp \Delta^< = \col.
\ee

Using the hermiticity condition $\Delta^{<\dagger} = -\Delta^<$ this equation
can be split into its hermitian and antihermitian parts
\bea
\lp k^2 + \frac{1}{4} \partial_z^2 \rp \Delta^< - \frac 12 \acom{\cm}{\Delta^<}
- \frac{i}{4} \com{\cm^\prime}{ \partial_{k_z} \Delta^<} = 0
\label{CE}
 \\
 k_z \partial_z \Delta^< + \frac i2\com{\cm}{\Delta^<}
   - \frac{1}{4} \acom{\cm^\prime}{ \partial_{k_z} \Delta^<} = \col
\label{KE}
\eea
where $\com{\cdot}{\cdot}$ and $\acom{\cdot}{\cdot}$
denote commutators and anticommutators.
In the following we refer to these two equations as the constraint
and kinetic equation.

\subsection{Lowest Order Solution}

Let us first discuss equations~(\ref{CE}--\ref{KE})
for a two-dimensional mass matrix that is constant in space and time.
The mass matrix can be diagonalized by a unitary
transformation and the equation in this basis reads (${\cal M}^2_d$
denotes the diagonalized mass
matrix and $\Delta_d$ the corresponding Green function that is non-diagonal in general)
\bea
 \lp k^2 + \frac{1}{4} \partial_z^2 \rp \Delta_d^<
 - \frac 12 \acom{{\cal M}^2_d}{\Delta_d^<} = 0 \\
k_z \partial_z \Delta_d^< + \frac i2\com{{\cal M}^2_d}{\Delta_d^<} = \col_d
\,.
\label{scalar kinetic 0th order}
\eea
The question is, in which sense these equations can recover the
solution in thermal equilibrium (\ref{delta_eq}).
We expect that the Kubo-Martin-Schwinger (KMS) equilibrium condition is
then satisfied, such that $\col_d = 0$.
We can use the derivative of the second equation to obtain
\bea
  k^2 \Delta_d^< - \frac{1}{16k_z^2}
         \com{{\cal M}^2_d}{\com{{\cal M}^2_d}{\Delta_d^<}}
- \frac{1}{2}\acom{{\cal M}^2_d}{\Delta_d^<} = 0
\label{con_0}
 \\
k_z \partial_z \Delta_d^<
       + \frac i2\com{{\cal M}^2_d}{\Delta_d^<}
  = 0
\,.
 \label{kin_0}
\eea
Note that, upon the identification, $m^\dagger m$ ($mm^\dagger$) with
${\cal M}^2$, these equations become identical
to the leading order equations obtained for the chiral fermionic
distribution functions $g_R$ ($g_L$)
in Ref.~\cite{fermions}.
The constraint equation (\ref{con_0})
 is algebraic, and it determines the spectrum of
the quasiparticles in the plasma.
At this point it is helpful to introduce two projection operators
\be
P^T X =\frac{1}{\Lambda^2} \com{{\cal M}^2_d}
                           {\com{{\cal M}^2_d}{X}}, \quad P^D = 1 - P^T,
\ee
where $\Lambda:=\sqrt{\textrm{Tr} \cm - 4 \textrm{ Det} \cm}=
\textrm{Tr} \, (\sigma_3 {\cal M}^2_d)$ denotes the difference
 of the eigenvalues of
${\cal M}^2_d$ and $\sigma_i$ ($i=1,2,3$) are the Pauli matrices.
The properties of the projection operators
\be
(P^T)^2 = P^T, \quad (P^D)^2 = P^D, \quad P^T+P^D=1
\ee
can be easily checked.

In the mass eigenbasis  $P^T \Delta_d^<$ corresponds to the complex off-diagonal entries,
while $P^D \Delta_d^<$ corresponds to the two real diagonal entries. If we split $\Delta^<_d$
in its transverse and diagonal parts
$\Delta_d^T:=P^T\Delta_d^<,\quad \Delta_d^D:=P^D\Delta_d^<$ and using the relations
\bea
\acom{Y^D}{X^D} = 2 Y^D \, X^D, &\quad& \acom{Y^D}{X^T} = (\textrm{Tr }Y) X^T, \\
P^D {\cal M}^2_d ={\cal M}^2_d, &\quad& P^T {\cal M}^2_d = 0
\,,
\eea
the constraint equations~(\ref{con_0})
for the diagonal and transverse parts of $\Delta_d^<$ decouple
\bea
\lp k^2 - {\cal M}^2_d \rp  \Delta_d^D &=& 0, \\
\lp k^2 - \frac{\Lambda^2}{16k_z^2} - \frac{1}{2}\textrm{Tr }{\cal M}^2_d
 \rp \Delta_d^T &=& 0
\,.
\label{on shell constraints}
\eea
Both diagonal and transverse constraint equation are algebraic, and thus
the solutions are given by the appropriate $\delta$-functions, which
represent sharp on-shell projections.
The diagonal shell is given by the standard dispersion relation,
whose frequencies are, $k_0^2 \equiv \omega_i^2
 = \vec k^2 +m_i^2$, where
$m_i^2$ are the eigenvalues of $\cm$. The
transverse parts fulfill a different on-shell condition,
which can be easily obtained from~(\ref{on shell constraints}).
 Note that these on-shell conditions are the same as
the ones found in~\cite{fermions} by solving the leading order
constraint equations for fermions.

The kinetic equation (\ref{kin_0}) reveals another difference between
 diagonal and transverse parts. The kinetic equations read
\bea
k_z \partial_z \Delta_d^D &=& 0, \\
k_z \partial_z \Delta_d^T
  + \frac i2\com{{\cal M}^2_d}{\Delta_d^T} &=& 0
\,.
\eea
The diagonal parts are constant in space and time,
while the transverse parts rotate
in flavour space with the frequency $\sim {\Lambda}/{k_z}$.

In the equilibrium solution (\ref{delta_eq}) the transverse entries
vanish everywhere, but it is clear that this oscillation dominates
the dynamics of the transverse parts as soon as they are sourced
by higher order contributions in the gradient expansion.

 Alternatively, oscillations can be induced by the initial conditions.
This is, for example, the case in neutrino oscillations. Neutrinos are
namely created as flavor eigenstates,
and hence, from the point of view of the mass eigenbasis,
a mixture of diagonal and transverse states.
Since in most environments the damping of neutrinos is very small,
neutrino oscillations persist for a long time.

\subsection{First order solution and CP violation}

Let us consider again the Kadanoff-Baym equations~(\ref{CE}--\ref{KE})
to first order in gradients.
In the last section we saw that in lowest order the spectrum can be
separated into the diagonal and transverse contributions.
One can show that in the first order system~(\ref{CE}--\ref{KE}) however,
the different quasiparticles start to mix and the spectral
functions acquire a finite width.
This is  reflected in the fact that, at first order in gradients,
the constraint equation is not any more algebraic.

Fortunately we do not need any information about the spectrum
to solve the kinetic equation~(\ref{KE}),
since it does not explicitly contain any $k_0$ dependence.
When transformed into the mass eigenbasis, the kinetic equation reads
\bea
k_z \partial_z \Delta_d^<  + k_z \com{\Sigma}{\Delta_d^<}
   + \frac i2\com{{\cal M}^2_d}{\Delta_d^<}
   -  \frac{1}{4} \acom{{{\cal M}^2_d}^\prime
   + \com{\Sigma}{{\cal M}^2_d}}{ \partial_{k_z} \Delta_d^<} = \col_d
\label{kin_1}
\eea
with
\begin{equation}
\Sigma= U^\dagger U^\prime
\label{Sigma}
\end{equation}
and the matrix $U(z)$ diagonalizes $\cm$,
 ${\cal M}^2_d = U^\dagger \cm U$.

The next step is to determine the CP violating contributions to the particle densities.
By definition the CP conjugation acts as
\bea
&\Delta_d^{\CP}(X,k)\,\equiv
\, \CP \,\, \Delta_d(X,k) \,\, \CP \,=\,  \Delta_d^*(\bar X,-\bar k), &\\
& \quad \bar X^\mu = (X_0, -X_i), \quad\bar k^\mu = (k_0, -k_i)\,.&
\eea
This transformation is in our equation (\ref{kin_1}) equivalent to
\be
U \rightarrow U^*, \quad \Sigma \rightarrow \Sigma^*
\,.
\ee
Now suppose that as in the chargino case our particles do not directly
couple to the sphaleron process. Then the CP-violating particle density
has to be communicated to the other species via interactions.
Therefore we are rather interested in the CP violating densities in the
diagonal matrix elements of the Green function in the interaction eigenbasis.
These are given by
\bea
\textrm{Tr } [ \Delta^< - \CP\, \Delta^< \CP ] &=&
\textrm{Tr } \big[
                  U\Delta_d^< U^\dagger - U^* \, \Delta^{<\CP}_d U^{\dagger*}
             \big]
%\nn\\
%&=& \textrm{Tr}\big[
%                    U \Delta_d^< U^\dagger - U \, \Delta^{<\CP*}_d U^\dagger
%               \big]
\nn\\
&=& \textrm{Tr }\big[
                    U ( \Delta_d^< - \Delta^{<\CP*}_d ) U^\dagger
                \big]
\eea
and
\bea
\textrm{Tr } \big[ \sigma_3 \Delta^< - \sigma_3\CP\, \Delta^< \CP
             \big]
     &=&
\textrm{Tr } \big[ \sigma_3 U \Delta_d^< U^\dagger - \sigma_3 U^* \,
                    \Delta^{<\CP}_d U^{\dagger*}
             \big]
 \nn\\
%&=& \textrm{Tr }\big[
%                     \sigma_3 U \Delta_d^< U^\dagger -  \sigma_3 U
%                      \, \Delta^{<\CP*}_d U^\dagger
%                \big]
%\nn\\
&=& \textrm{Tr }\big[
                     \sigma_3 U ( \Delta_d^< - \Delta^{<\CP*}_d ) U^\dagger
                \big],
\eea
where the latter equality in both cases follows from the fact that
$\Delta_d^{<\CP}$ is hermitian.
Henceforth we consider in the mass eigenbasis the equation for
$\Delta^{<\Q}:=\Delta^{<\CP*}$.
This Q-conjugation coincides with CP-conjugation on the diagonal,
but it is in addition basis independent, since it commutes with the
diagonalization matrix.
This fact was already used in~\cite{fermions} to identify
CP-violating quantities for mixing fermions before the Green function
was transformed to the interaction eigenbasis.

The equation for $\Delta^{<\Q}$  is given by (notice that $\Sigma$ is antihermitian)
\bea
k_z \partial_z \Delta_d^{<\Q}  + k_z \com{\Sigma}{\Delta_d^{<\Q}}
 - \frac i2\com{{\cal M}^2_d}{\Delta_d^{<\Q}}
 - \frac{1}{4} \Big\{{{\cal M}^2_d}^\prime
    + \com{\Sigma}{{\cal M}^2_d}, \partial_{k_z} \Delta_d^{<\Q}\Big\}
 = \col_d
\,.
\eea
The only change with respect to the original equation of $\Delta_d^<$
is a sign-change in the oscillation term
$\big[{\cal M}^2_d,\Delta_d^{<Q}\big]$.
If we include higher order terms in the
gradient expansion additional Q breaking terms will appear.
  Since in leading order
CP violation is based on the oscillation effect, one has to solve only the
equation of the transverse parts and its Q conjugate. Collecting terms,
that are at most first order in gradients
(deviations from equilibrium $\delta\Delta_d=\Delta_d-\Delta_{\rm eq}$,
${{\cal M}^2_d}^\prime$ and $\Sigma$ are
counted as of order one in the gradient expansion) we get
for the transverse deviations,
\begin{eqnarray}
k_z \partial_z \delta\Delta_d^T
 +\; \frac i2\com{{\cal M}^2_d}{\delta\Delta_d^T} \;- \col_d
   &=& \cs_d \nn \\
k_z \partial_z \delta\Delta_d^{T\Q}
 - \frac i2\com{{\cal M}^2_d}{\delta\Delta_d^{T\Q}} - \col_d
   &=& \cs_d
\,,
\label{calc_equ}
\end{eqnarray}
with the source term
\bea
\cs_d = - k_z \com{\Sigma}{\Delta_{\rm eq}^<}^T
        + \frac{1}{4} \acom{{{\cal M}^2_d}^\prime
            + \com{\Sigma}{{\cal M}^2_d}}{ \partial_{k_z} \Delta_{\rm eq}^<}^T
\,.
\label{bos_src}
\eea
This can be solved numerically using an {\it Ansatz} for a flow solution
as described in Ref.~\cite{fermions}.

Since $\Delta_d^{<CP}$ and $\Delta_d^{<Q}$
differ only by transposition, this calculation in addition shows that the
diagonal entries in the mass eigenbasis will be CP even up to first order
in gradients.

\section{The Damping Term \label{sec_damp}}

If we solve equations (\ref{calc_equ}) without the collision term, we will have problems to ensure
that our solution will be close to thermal equilibrium
on both sides at a large distance from the wall.
This problem can be solved by introducing a damping term,
 that corresponds to statistical effects
due to the interaction of the particles with the heat bath.
 In the rest frame of the plasma, the damping should take place
in the positive time like direction as {\it e.g.} in the equation
\bea
k\cdot \partial_X \Delta + k_0 \Gamma \Delta = \cs
\eea
In the wall frame this leads to
$\col_d = \gamma_{v_w}( k_0 - v_w k_z )\, \Gamma\, \Delta$ with the wall
velocity $v_w$. For $\Gamma$ a reasonable choice is
$\Gamma = \alpha T_c$, where $\alpha$ denotes
the coupling strength of the dominant interaction of the species and $T_c$ is
the temperature of the plasma during the phase transition.

However, by introducing a term that breaks time reversal invariance,
we run the risk of breaking CP explicitly by introducing
new artificial CP-violating sources.
We illustrate this by the following simple example.
Assume that a quantity $W$, which denotes a CP-violating deviation
from equilibrium, fulfills the equation
\bea
\partial_z \, W = \exp(-z^2) \, n_{BE}(\sqrt{k_z^2+m^2}) \equiv \cs(z,k_z)
\eea
and we are interested in $\int dk_z \, W(z,k_z)$.

To solve this equation, we can use the Green function method with the boundary
condition, such that $W$ vanishes in the unbroken phase ($z \to -\infty$),
where the wall has not yet influenced the plasma. Then
\be
W(z,k_z) = \int dz^\prime \,\, g(z,z^\prime) \,\, \cs(z^\prime,k_z)
 \label{toy_without}
\,,
\ee
with the Green function
\be
g(z,z^\prime)=\theta(z-z^\prime)
\ee
and $\int dk \, W(z,k_z)$ can be determined.

Since the solution does not vanish in the broken phase ($z \to +\infty$), we
introduce a phenomenological damping term that breaks time invariance
and our choice could be in analogy to the considerations above
\bea
\partial_z W + \frac{k_0}{k_z} \Gamma \,\, W = \cs(z,k_z)
 \label{toy_with}
\eea
The corresponding Green function is
\bea
g(z,y) = \left\{
\begin{matrix}
k_0/k_z > 0: & \exp\Big(-({k_0}/{k_z}) \Gamma (z-z^\prime)\Big)
                         \, \theta(z-z^\prime)
 \\
k_0/k_z < 0: & -\exp\Big(({k_0}/{k_z}) \Gamma (z-z^\prime)\Big)
                        \, \theta(z^\prime-z)
\end{matrix}
\right.
\eea
and yields the desired result. On the other hand, if the source is
odd in $k_z$, the picture changes. The equation (\ref{toy_without})
yields a solution, that is odd in $k_z$ and $\int dk_z \, W(z,k_z)$ vanishes,
while the solution of (\ref{toy_with}) gives a non-vanishing result even
after integration over $k_z$.

The same effect can be seen in the kinetic equation (\ref{calc_equ}). Without
the damping term, the result will be odd in $k_z$, such that only the three
component of the particle current
\be
j_\nu(X_\mu) = 2i \int_{k_0>0} \frac{d^4k}{(2\pi)^4}
                    \,  k_\nu \, \Delta(X_\mu, k_\mu).
\ee
is sourced. This is expected, since if this current is Lorentz boosted into
the rest frame of the plasma, the CP-violating particle density
 $j_0^{plasma-frame} = \gamma_{v_w}\, v_w \, j_3^{wall-frame}$ vanishes in
the static wall limit, $v_w=0$.

After the damping term is introduced,
$j_0^{plasma-frame}$ is sourced even in the case
of a static wall profile, which is clearly an unphysical result
for a CP-violating quantity. Notice that
this source persists even in the limit $\Gamma_h \to 0$. In the following we
keep only the source terms, which are not induced by the damping term.

\section{Transport Equations for Mixing Charginos}
\label{sec_ferm}

In this section we recall the fermionic
transport equations derived in \cite{fermions}.
Due to the additional spinor structure of the Green function, we have to
solve two equations for the left-handed and right-handed densities separately.
In addition, the Green functions have a spin quantum number $s$.
As in the bosonic case, only the transverse parts oscillate and
contribute to the CP-violating (or better Q-violating) densities.
In the mass eigenbasis the equations for
$\delta g_{Rd}^{Ts}$ and $\delta g_{Ld}^{Ts}$ read
(see Eq.~(78) in Ref.~\cite{fermions})
\bea
k_z \partial_z \delta g_{Rd}^{Ts}
  + \frac{i}{2} \com{m^2_d}{\delta g_{Rd}^{Ts}}
+ k_0 \Gamma_h \delta g_{Rd}^{Ts} &=& \cs_R^s \\
k_z \partial_z \delta g_{Ld}^{Ts}
  + \frac{i}{2} \com{m_d^2}{\delta g_{Ld}^{Ts}}
+ k_0 \Gamma_h \delta g_{Ld}^{Ts} &=& \cs_L^s
\,,
\eea
with the spin-dependent part of the sources
\bea
\cs_R^s &=&
- s\frac{k_z^2}{\tilde k_0} \com{ V V^{\dagger\prime}}{ g_{0,\rm eq}}
- \frac{s}{4\tilde k_0}\com{V(m^{\dagger\prime}m
   - m^\dagger m^\prime)V^\dagger}{g_{0,\rm eq}}
+ \frac{sk_z}{4\tilde k_0}
      \acom{V (m^\dagger m)^\prime V^\dagger}{g_{0,\rm eq}}^T
\nn \\
\cs_L^s &=&
s\frac{k_z^2}{\tilde k_0} \com{ U U^{\dagger\prime}}{ g_{0,\rm eq}}
+ \frac{s}{4\tilde k_0}\com{U(m^{\prime}m^\dagger
   - m m^{\dagger\prime})U^\dagger}{g_{0,\rm eq}}
- \frac{sk_z}{4\tilde k_0}
       \acom{U(m m^\dagger)^\prime U^\dagger}{g_{0,\rm eq}}^T
\!\!.
\quad
\label{ferm_srces}
\eea
The function $g_{0,\rm eq}$
 denotes the $\gamma_0$ coefficient of the Green function
in thermal equilibrium and mass eigenbasis
\be
g_{0,\rm eq}= 2\pi |k_0| \delta(k^2 - m_d^2) f_{FD}
, \quad f_{FD}=\frac{1}{e^{\beta k_0}+1}.
\label{g0_eq}
\ee
The chargino mass matrix $m$ is given by
\begin{equation}
  m = \left(\begin{array}{cc}
                    M_2 & gH_2 \cr
                    gH_1  & \mu_c \cr
      \end{array}\right)
\end{equation}
and diagonalized by the biunitary transformation
\begin{equation}
  m_d = UmV^\dagger
\,,
\end{equation}
where $\mu_c$ and $M_2$ are the soft supersymmetry breaking parameters.

To compare the result from these equations with the work 
\cite{Madrid_group,Madrid_group_imp}
it is helpful to examine the contributions of the different sources in
the local approximation, $\Gamma_h \to \infty$, in which
diffusion transport is neglected. In this case, the resulting
CP-violating vector and axial vector particle currents behave as
\begin{eqnarray}
&& \hskip -2cm {\rm Tr}\big(\sigma^3 j_{5\,\mu}\big) = \cs^a_\mu, \quad
{\rm Tr}\big(\sigma^3 j_\mu \big) = \cs^b_\mu + \cs^c_\mu
\nn \\
\cs^a_\mu &=& 2T_c^{-4}\Im(M_2\mu_c) (|M_2|^2 - |\mu_c|^2)
             \partial_\mu \big(u_1 u_2\big)\eta_{(0)}^3
\nn \\
\cs^b_\mu &=&  2T_c^{-4}\Im(M_2\mu_c) (u_1^2-u_2^2)
       \partial_\mu(u_1 u_2)\eta_{(0)}^3
\nn \\
\cs^c_\mu  &=& -2T_c^{-2}
      \Im(M_2\mu_c)\big( u_2\partial_\mu u_1 - u_1\partial_\mu u_2\big)
                   \big(\eta_{(0)}^0 + 4\eta_{(2)}^3\big)
\,,
\label{sources_abc}
\end{eqnarray}
where $u_{1,2} = g|H_{1,2}|$, and $\eta_{(0)}^0$,
$\eta_{(0)}^3$ and $\eta_{(2)}^3$
are integrals derived in~\cite{fermions},
\begin{eqnarray}
 \eta_{(n)1,2} &=& T_c^{2-n}\int_{k_0>0}\frac{d^4 k}{(2\pi)^3}
                  \frac{k_0}{\tilde k_0}k_z^n
                  \frac{n(k^\mu,m^2_{1,2})}{k_0^2\Gamma_h^2+(\Lambda/2)^2}
                  \delta(k^2 - m_{1,2}^2)
\nonumber\\
  \eta_{(n)}^0 &=& \frac 12 \big(\eta_{(n)1}+\eta_{(n)2}\big)
\,,\qquad
  \eta_{(n)}^3 = \frac{T_c^2}{2\Lambda} \big(\eta_{(n)1}-\eta_{(n)2}\big)
\,,
\end{eqnarray}
and $n=n(k^\mu,m^2_{1,2})$ denotes the distribution function.
The contributions $\cs_a$ and $\cs_b$ result from the
first term in the sources (\ref{ferm_srces}), while the term $\cs_c$
results from the second and third terms in the sources (\ref{ferm_srces}).

Comparing with Eq. (3.13) of \cite{Madrid_group_imp}, we see that
in local approximation our sources agree in the characteristics
of the $z$-dependence, but show different structure in momentum space.

 To facilitate a comparison with the work on semiclassical force
baryogenesis of Cline, Joyce and
Kainulainen~\cite{ClineJoyceKainulainen:2000},
we quote the dominating local source at the second order in gradients
in the plasma frame~\cite{fermions,dickwerk},
\bea \textrm{Tr} \lp\mathbbm{1} \,j^{(2)}_{5,0} \rp
  \equiv \cs^d_0 =  2 \, v_w\, T_c^{-4}\Im(M_2\mu_c)
                  \big(u_2\partial^2_z u_1 + u_1\partial^2_z u_2\big)
                   \zeta_{(0)}^3
\,, \label{source d}
 \eea
where $\zeta_{(0)}^3 =\eta_{(0)}^3|_{\Lambda\rightarrow 0}$.
This source corresponds to the CP-violating shift in the
dispersion relation and dominates if the mass spectrum
in the chargino sector is far from degeneracy ($\Lambda \to \infty$)
and in the limit of a small damping.
It contributes in contrast to the first order terms to the trace of
the chargino current.

\section{Diffusion Equations\label{sec_diff}}

Using our formalism, we can deduce the CP-violating particle densities in
the chargino sector. To evaluate the baryon asymmetry in the broken phase,
we need to compute the density of left-handed quarks and leptons $n_L$ in
front of the wall. These densities couple to the weak sphaleron and
produce a net baryon number.

To determine how the CP-violating currents are transported from
the charginos to the left-handed quarks and leptons we use a
system of coupled diffusion equations as derived in
\cite{Huet_Nelson}, and later adapted in~\cite{Carena:1997gx,Madrid_group}
and~\cite{ClineJoyceKainulainen:2000}. The diffusion equations are
\bea
v_w \, n_Q^\prime &=& D_q \, n^{\prime\prime}_Q
- \Gamma_Y \left[ \frac{n_Q}{k_Q} - \frac{n_T}{k_T} - \frac{n_H+n_h}{k_H} \right]
- \Gamma_m \left[ \frac{n_Q}{k_Q} - \frac{n_T}{k_T}  \right] \nn \\
&& -6 \, \Gamma_{ss} \left[ 2\, \frac{n_Q}{k_Q} - \frac{n_T}{k_T} + 9\, \frac{n_Q+n_T}{k_B} \right]
\label{diff_Q} \\
v_w \, n_T^\prime &=& D_q \, n^{\prime\prime}_T
+ \Gamma_Y \left[ \frac{n_Q}{k_Q} - \frac{n_T}{k_T} - \frac{n_H+n_h}{k_H} \right]
+ \Gamma_m \left[ \frac{n_Q}{k_Q} - \frac{n_T}{k_T}  \right] \nn \\
&& +3 \, \Gamma_{ss} \left[ 2\, \frac{n_Q}{k_Q} - \frac{n_T}{k_T} + 9\, \frac{n_Q+n_T}{k_B} \right]
\label{diff_T} \\
v_w \, n_H^\prime &=& D_h \, n^{\prime\prime}_H
+ \Gamma_Y \left[ \frac{n_Q}{k_Q} - \frac{n_T}{k_T} - \frac{n_H+n_h}{k_H} \right]
- \Gamma_h \frac{n_H}{k_H}
\label{diff_H} \\
v_w \, n_h^\prime &=& D_h \, n^{\prime\prime}_h + \Gamma_Y \left[
\frac{n_Q}{k_Q} - \frac{n_T}{k_T} - \frac{n_H+n_h}{k_H} \right] -
(\Gamma_h+4\Gamma_\mu) \frac{n_h}{k_H} \,,
 \label{diff_h} \eea
where $n_T$ denotes the density of the left-handed top and stop particles,
$n_Q$ the remaining left-handed quarks and squarks and $n_H$ and $n_h$ the
sum and difference of the two Higgsino densities $n_{H_1}$ and $n_{H_2}$.
The quantities $k_i$ are statistical factors defined by
$n_i = k_i \, \mu_i \frac{T_c^2}{6}$ ($\mu_i$ denotes
the chemical potential of species $i$).
For light, weakly interacting particles $k_i \approx 2$ (bosons) or
$k_i \approx 1$ (fermions), while for particles much heavier than $T_c$, $k_i$ is
exponentially small. We use the values
\be k_Q \approx 6,\quad  k_T \approx 9,\quad k_B \approx 3,\quad
k_H \approx 12 \label{k_param} \ee 
corresponding to the light stop scenario \cite{Huet_Nelson} and the diffusion
constants are~\cite{JoyceProkopecTurok:1994-I}
\be
D_q \sim 6/T_c,\quad D_h \sim 110/T_c.
\ee
For the particle number changing rates we take
\cite{JoyceProkopecTurok:1994-I,JoyceProkopecTurok:1995,Huet_Nelson},
\bea
\Gamma_y \approx \frac{1}{10} T_c, \quad
\Gamma_m \approx \frac{1}{10} T_c, \quad
\Gamma_h \approx \frac{1}{20} T_c, \quad
\Gamma_\mu \approx \frac{1}{10} T_c
\eea
and for the sphaleron rates \cite{sphaleron}
\bea
\Gamma_{ss} \approx 1.5 \times 10^{-2}\, T_c, \quad
\Gamma_{ws} \approx 6.0 \times 10^{-6}\, T_c.
\eea
The diffusion equations~(\ref{diff_Q}--\ref{diff_h}) are derived
under the assumptions~\cite{ClineJoyceKainulainen:2000} that (a)
the supergauge interactions, which are of the weak strength, are
in equilibrium; (b) the chargino asymmetry gets transported to the
quark sector via the strong top Yukawa interactions, while the
wino asymmetry does not contribute; (c) the gaugino helicity-flip
interactions are in equilibrium, implying that the chemical
potentials for particles and their supersymmetric partners are
equal. These approximations imply that the main channel for baryon
production is the conversion of the chargino asymmetry into the
top sector, which then bias electroweak sphalerons. The accuracy
of these approximations will be addressed elsewhere.

The solution of Eqs.~(\ref{diff_Q}--\ref{diff_h}) is performed in
several steps. First we use the transport equations in the
chargino sector as described in \cite{fermions} to determine $n_H$
and $n_h$. The result is used as an input in the equations
(\ref{diff_Q}) and (\ref{diff_T}). From these equations the
left-handed particle density $n_L = 5\, n_Q + 4 \, n_T$ can be
determined and used as a source for the weak sphaleron process as
described in \cite{Madrid_group} (see also
Ref.~\cite{ClineKainulainen:2000}). The net baryon density is
given by
%\
\be
n_B = - \frac{3\, \Gamma_{ws}}{v_w} \int_{-\infty}^0 dz \,\, n_L(z)
\exp\left( z\frac{15\Gamma_{ws}}{4\,v_w} \right)
\ee
and finally the baryon-to-entropy ratio is determined via
\be
\eta = \frac{n_B}{s}, \quad s = \frac{2\pi^2}{45}
   g_{\rm eff} T_c^3 \approx 51.1\,\, T_c^3.
\ee
To check, whether our solution of the diffusion equation is consistent, we used the
densities $n_Q$ and $n_T$ as input for the equations (\ref{diff_H}) and (\ref{diff_h}).
The resulting deviations in the Higgsino densities never exceed $5\%$ of the
original densities. This is due to the fact that the Higgsino diffusion constant $D_h$
is rather large and that the oscillation partially suppresses
an efficient transport of the quarks and squarks. In this light the equations of
the Higgsinos decouple, since the oscillation provides the shortest time-scale.

Note that in the work~\cite{Giudice:1993bb} a suppression was found for the 
parameters of the Standard Model ($k_T \approx 3$ in Eq. (\ref{k_param})). 
As explained, for the mixing sources we consider here, 
the oscillation effectively decouples the dynamics of the charginos from the quarks/squarks. 
This allows us to neglect the backreactions from the quarks/squarks and 
leads to the absence of the suppression for $k_T \approx 3$.
If the oscillation is not the shortest time-scale, {\it i.e.} for 
$|\mu_c - M_2|< 5$ GeV, the backreactions become large and our 
approach does not reproduce the suppression 
of~\cite{Giudice:1993bb} and would indeed over-estimate the result. 
In the following we will employ the parameters of Eq. (\ref{k_param})
where this suppression mechanism is already ineffective. 

\section{Numerical Results\label{sec_num}}

\begin{figure}[htbp]
\begin{center}
\epsfig{file=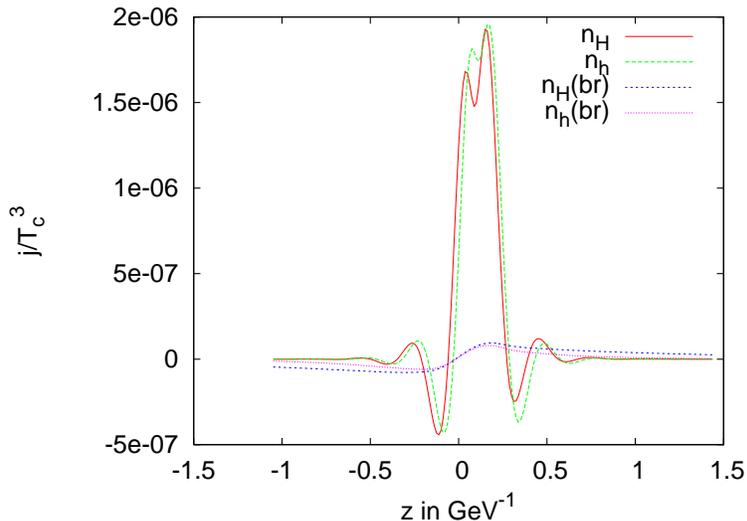, width=4.0 in}
\end{center}
\vskip -0.2in
\lbfig{plot_br}
\caption[Fig:br]{%
\small
The original Higgsino densities and the corresponding back-reactions.
The parameters of the plot are
$\mu_c=200$ GeV, $M_2=180$ GeV, $m_A=200$ GeV
}
\end{figure}
In this section we will present numerical results of the transport and
diffusion equations.
The Higgs {\it vevs} and the $\beta$ angle are parametrized by
\be
H_1(z) = H(z) \sin(\beta(z)), \quad
H_2(z) = H(z) \cos(\beta(z))
\ee
and
\bea
H(z) = \frac{1}{2} v(T)
  \lp 1- \tanh \lp \alpha \lp 1- \frac{2z}{L_w}\rp\rp \rp,
\label{H}
\\
\beta(z) = \beta_\infty - \frac{1}{2} \Delta\beta
\lp 1+ \tanh \lp \alpha \lp 1- \frac{2z}{L_w}\rp\rp \rp.
\label{beta}
\eea
The parameters used are
\begin{equation}
T_c=110~{\rm GeV} ,\quad
 v(T)=175~{\rm GeV},\quad
\alpha=\frac{3}{2},\quad \tan(\beta_\infty)=10,\quad
L_w=20/T_c
\end{equation}
and the complex phase is chosen maximally
\begin{equation}
\Im(M_2 \mu_c)=|M_2 \mu_c| \,.
\end{equation}
We have checked, with the help of a program developed by the authors 
of Refs.~\cite{Madrid_group,Carena:1997ki}, that the values for $v(T)$
compatible with present Higgs bounds 
typically lie in the range 165-185~GeV. The exact value depends on 
parameters of the Higgs and squark sectors which affect our results only 
through this expectation value. We therefore have fixed the value of $v(T)$ 
to its zero temperature result. The uncertainty arising from our choice is 
below ten percent.

The values of $\Delta\beta$ are deduced from \cite{Moreno_Seco_PT}
for the different values of $m_A$. The wall velocity is taken to
be $v_w=0.05$ and the transport equations are evaluated using the
fluid {\it Ansatz} for the first six momenta. The parameters of
the diffusion equations are given in the last section.

The plot Fig.~\ref{plot_br} supports the claim that, within our
approximations and for our choice of parameters, the back-reaction of
left-handed quarks and squarks, $n_Q$, $n_T$, on the charginos can be
neglected. The amplitude of the Higgsino densities coming from the
back-reaction is always smaller than $3\%$ and never leads to
corrections of the baryon-to-entropy ratio larger than $5\%$.

\begin{figure}[htbp]
\begin{center}
\epsfig{file=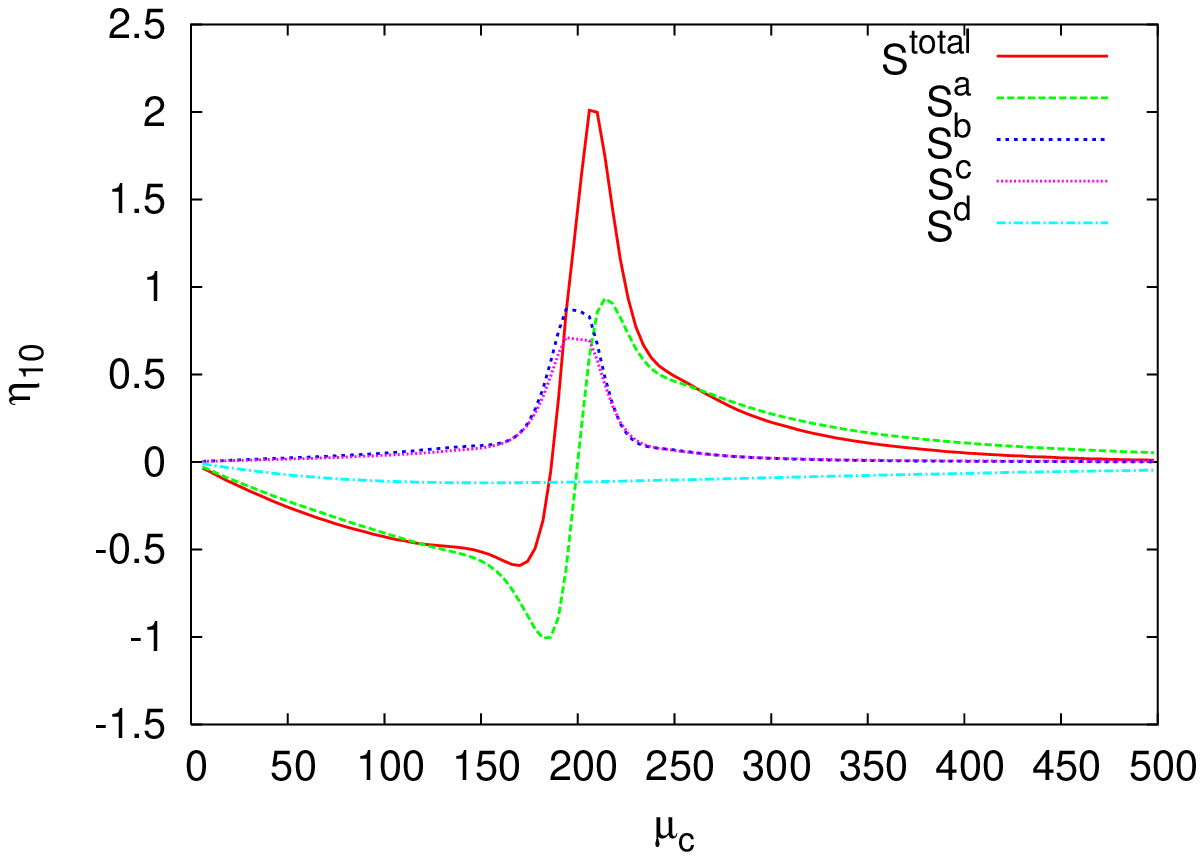, width=3.2in,height=3.2in}
\epsfig{file=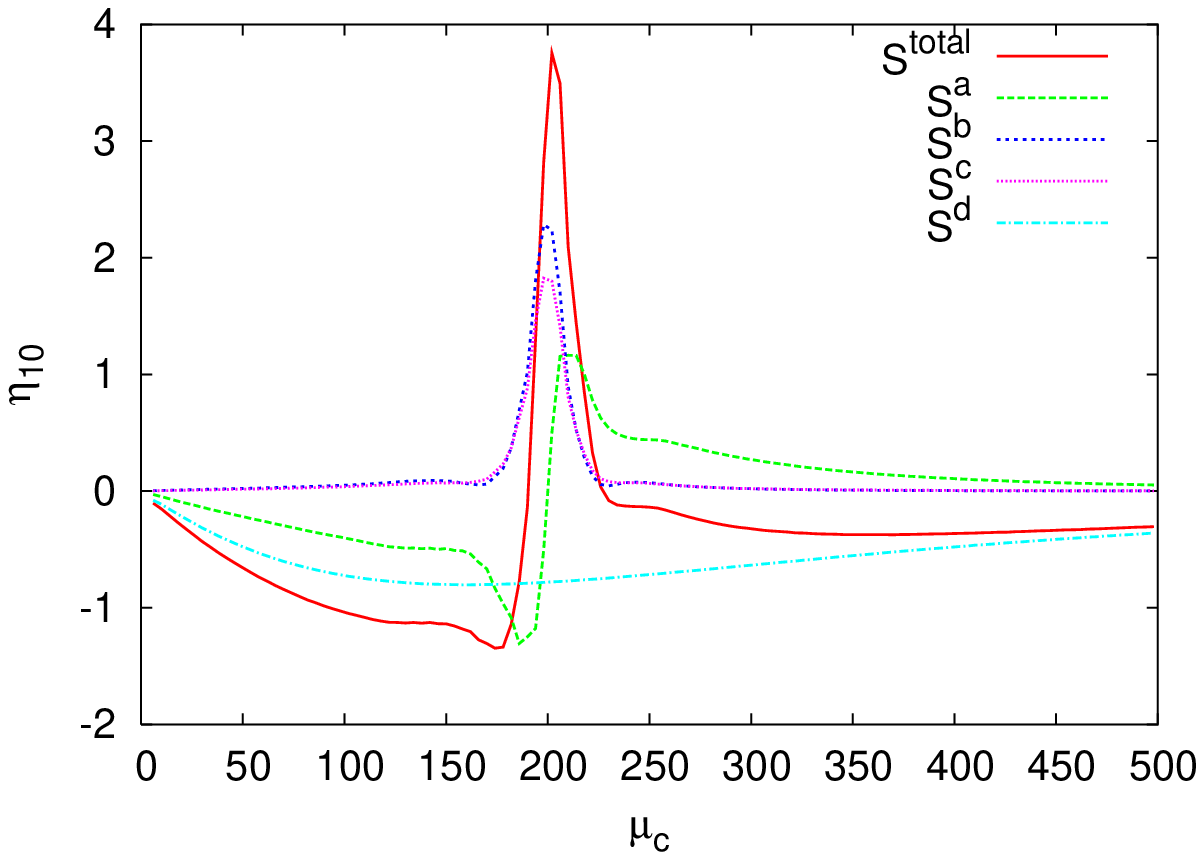,width=3.2in,height=3.2in}
\end{center}
\vskip -0.2in \lbfig{plot_source}
\caption[Fig:ma_2]{%
\small This plot shows the first and second order sources as a
function of $\mu_c$ with $M_2=200$~GeV. The plot on the left are
the sources with the damping, $\Gamma=\alpha_wT_c$, while on the
right plot, $\Gamma=0.25\alpha_wT_c$. }
\end{figure}
In Fig.~\ref{plot_source} we plot the first order sources ${\cal
S}_\mu^a$,  ${\cal S}_\mu^b$, ${\cal S}_\mu^c$ and the second
order source (semiclassical force) ${\cal S}^d$, as defined in
Eqs.~(\ref{sources_abc}--\ref{source d}). The first order sources
are roughly of the same magnitude, and they peak when
$|\mu_c|\simeq |M_2|$, where they also switch the sign. The second
order source varies slowly with $|\mu_c|$ and tends to dominates
when the difference $|\mu_c| - |M_2|$ becomes large. Note that
when the damping is small, the first order sources become more
peaked around $|\mu_c|=|M_2|$, but the amplitude of the baryon
asymmetry does not significantly change. On the other hand, the
second order source is about an order of magnitude larger in the
right plot, implying that in the limit of a small damping the
second order source (semiclassical force) may result in a viable
baryogenesis. Since our damping term is phenomenological and
flavor blind, it would be premature to conclude that the second
order source cannot lead to a viable baryogenesis, until a more
quantitative analysis of the damping term is performed.

\begin{figure}[htbp]
\begin{center}
\epsfig{file=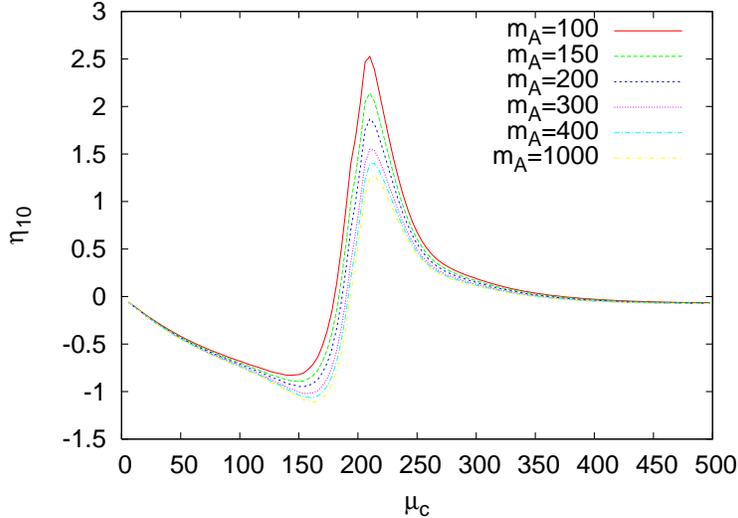, width=4.0 in}
\end{center}
\vskip -0.2in
\lbfig{plot_ma_2}
\caption[Fig:ma_2]{%
\small This plot shows $\eta_{10}=10^{10}\eta$ as a function of
$\mu_c$ with $M_2=200$ GeV and for several values of $m_A$ in GeV.
}
\end{figure}
The parameters chosen in Figs.~\ref{plot_ma_2}
and~\ref{plot_ma_1} are similar to the ones chosen in
\cite{Madrid_group_imp}, in order to facilitate comparison. In
plot Fig.~\ref{plot_ma_2} the parameters $m_A$ and $M_2$ are fixed
while $\mu_c$ is varied. The maximum is not exactly at $\mu_c=M_2$
as in \cite{Madrid_group_imp}, but rather close to $\mu_c \approx
M_2+20$~GeV. The reason for this difference is that in our case
all sources (\ref{sources_abc}) are of similar order, while in
\cite{Madrid_group_imp}, the baryon asymmetry is completely
dominated by a source term of form $\cs_\mu^c$ in
(\ref{sources_abc}) that is proportional to $\Delta\beta$ in the
parametrization (\ref{beta}) and hence suppressed for large values
of $m_A$ as shown in~\cite{Moreno_Seco_PT}.
Another difference is that our plot shows the suppression for
$\mu_c \gg M_2$ what is expected since in this case the
quasi-particles have highly separated on-shell conditions and
mixing should be suppressed.
We would like to emphasize that the peak around $\mu_c \approx
M_2+20$ GeV is due to this suppression and not due to a resonance in
the sources as it was in the 
publications~\cite{Riotto:1998zb,Madrid_group, Madrid_group_imp} 
and more recently in~\cite{Lee:2004we}. In the present work, the sources
show a resonance but the CP-violating densities do not
since they are generated by the oscillations
(see Eq. (\ref{calc_equ}) ) and contain near the degeneracy an additional
proportionality to the mass splitting $\Lambda$.
 In Fig.~\ref{plot_ma_1} the baryon
asymmetry is plotted near the maximal value $\mu_c \approx M_2+20$ GeV.
The maximum is reached near $\mu_c\approx 80$ GeV in contrast to
\cite{Madrid_group_imp} where the maximum was $\mu_c\approx 250$ GeV.
\begin{figure}[htbp]
\begin{center}
\epsfig{file=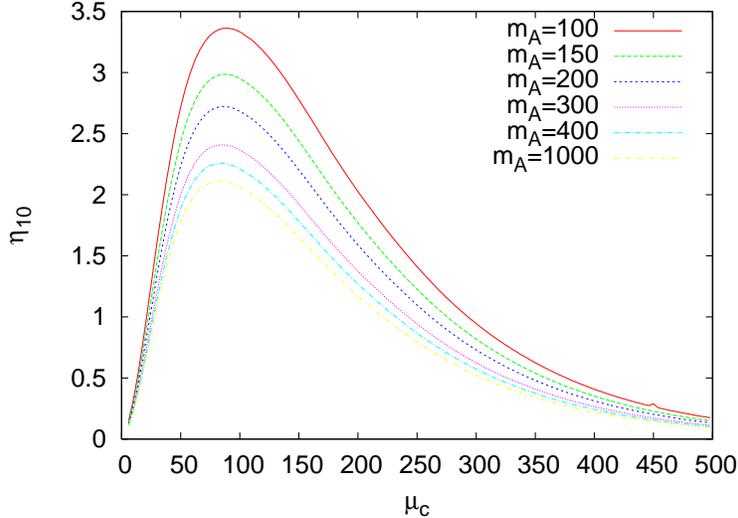, width=4.0 in}
\end{center}
\vskip -0.2in
\lbfig{plot_ma_1}
\caption[Fig:ma_1]{%
\small This plot shows $\eta_{10}=10^{10}\eta$ as a function of
$\mu_c$, $M_2=\mu_c - 20$ GeV and for several values of $m_A$ (in
GeV) . }
\end{figure}

Finally in Fig.~\ref{contour} two contour plots are shown with
regions in the $( M_2, \mu_c )$ parameter space for the baryon
asymmetry expressed in terms of $\eta_{10} \equiv 10^{10}\times
\eta$. In these units the observed value is close to unity,
$\eta_{10,\rm obs} = 0.8-0.9$. If $\eta_{10}>\eta_{10,\rm obs}$,
the observed value can be attained simply by adjusting the complex
phase, which is in our calculation chosen to be maximal. The two
plots correspond to the choices $m_A=200$ GeV and $m_A=400$ GeV.
\begin{figure}[htbp]
\begin{center}
\epsfig{file=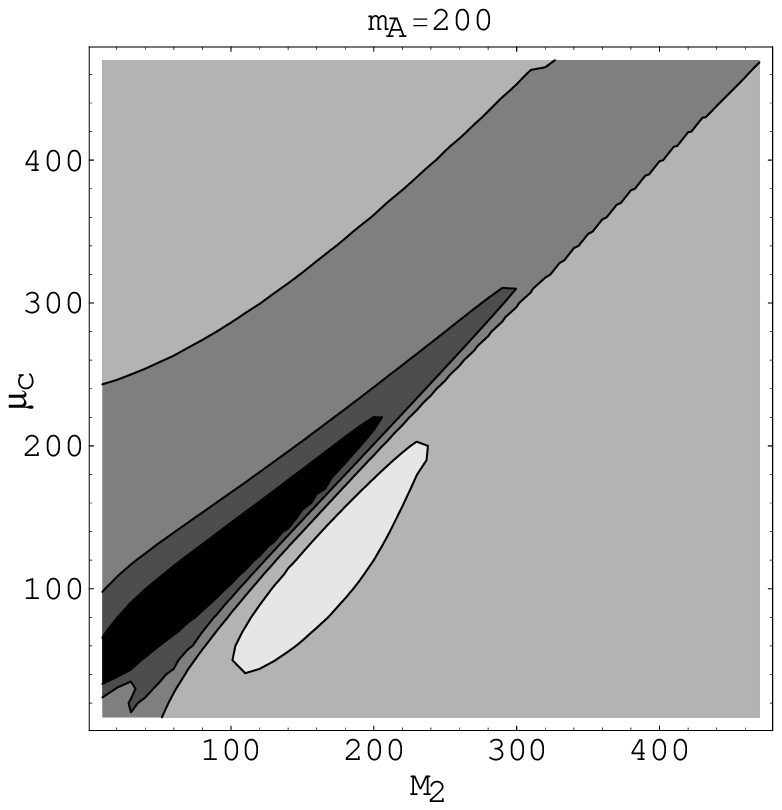, width=2.7 in}
\epsfig{file=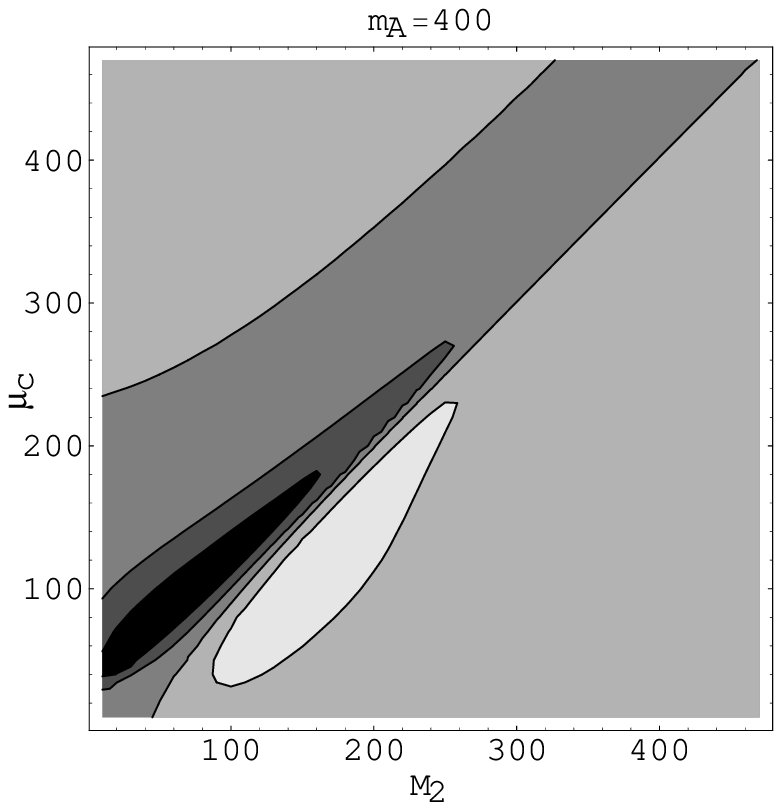, width=2.7 in}
\end{center}
\vskip -0.2in
\lbfig{contour}
\caption[Fig:contour]{%
\small The baryon-to-entropy ratio $\eta_{10}=10^{10}\times \eta$
in the $( M_2, \mu_c )$ parameter space from (0~GeV,0~GeV) to
(500~GeV,500~GeV). For the left plot the value $m_A=200$ GeV is
used, for the right plot $m_A=400$~GeV. The black region denotes
$\eta_{10}>1$, where baryogenesis is viable. The other four
regions are bordered by the values of $\eta_{10}$,
$\{-0.5,0,0.5,1\}$, beginning with the lightest color.}
\end{figure}

In the following we will comment on differences between 
the formalism used in this paper and the work~\cite{Madrid_group_imp}
that lead to the discrepancy in the numerical results~\footnote{
There are some differences between the results presented 
in~\cite{Madrid_group}, \cite{Madrid_group_imp} 
and~\cite{BalazsCarenaMenonMorrisseyWagner:2004}.
Nevertheless, these differences do not affect one of the main results 
of~\cite{Madrid_group, Madrid_group_imp,BalazsCarenaMenonMorrisseyWagner:2004},
that is, the presence of a resonance in the minus current for
values $|M_2|=|\mu|$. In the present paper we have found that this effect
is strongly suppressed by the oscillations induced by the off diagonal
terms.
}.
As already mentioned in a previous section the authors in~\cite{Madrid_group_imp}
work in the flavour basis and write classical Boltzmann equations
using CP-violating sources whereas in this work the sources 
appear genuinely in a basis independent set of quantum transport
equations. In this work the damping $\Gamma_h$ is primarily introduced 
to obtain consistent boundary conditions and it corresponds to the 
helicity flip rate $\Gamma_h$ in the diffusion equations of~\cite{Madrid_group}.
We have excluded $\Gamma_h$ dependent terms that violate CP and the limit 
$\Gamma_h \rightarrow 0$ is straightforward.

In addition to the damping $\Gamma_h$ a Breit-Wigner width 
$\Gamma_{\tilde t}$ was introduced in the chargino spectrum
in~\cite{Madrid_group}.
 We have checked in the simpler bosonic case that, 
for $\Gamma_{\tilde t} \to 0$, 
the present sources and those in~\cite{Madrid_group} 
are related in a simple way. A detailed 
discussion is presented in the Appendix.
%The use of the Wigner transform
%in this paper is most important for deriving transport
%equations from first principles and the direct calculation in 
%\cite{Madrid_group} just leads to the 
%same results for the sources.
Of course, the ambiguity related to the magnitude of the source in the
chargino diffusion equations remains in the formalism used in 
Refs.~\cite{Madrid_group,Madrid_group_imp},
where a phenomenological thermalization time $\tau$ or the classical Fick's law 
had to be used 
to incorporate the sources into the diffusion equations.
In our formalism the magnitude of the source is completely specified.

Furthermore, we have checked that the effect of the Breit-Wigner
broadening on our sources is small. This effect can be modeled by replacing
the $\delta$-function in (\ref{g0_eq}) by the corresponding Breit-Wigner form.
To account for the finite $\Gamma_{\tilde t}$ in the transport and not 
just in the sources is on the same level as a treatment of the collision 
terms in Refs.~\cite{Kainulainen:2002sw, dickwerk, Lee:2004we}
and it is outside the scope of this paper.
In principle, the collision term could as well yield additional CP-violating
sources, but a one-loop calculation~\cite{dickwerk} in a model theory
of chiral fermions Yukawa-coupled to scalars, indicates
that the collisional sources are phase space suppressed
with respect to our tree level sources.

\section{Conclusions and Discussion\label{sec_conc}}

In this work we obtained the baryon asymmetry of the universe
during the electroweak phase transition in the MSSM using
semiclassical transport equations derived in a first principle
approach from the Kadanoff-Baym (KB) equations in Ref.~\cite{fermions}.
When the KB equations are expanded in gradients
in the general case of mixing fermions,
the CP-violating deviations from equilibrium can be sourced 
by a space-time dependent Higgs background both at first and second order
in gradients. The first order effects, 
which occur only in the presence of fermion mixing, 
have been consistently determined including oscillations that are
crucial for the dynamics of the CP-violating densities. 
The second order effects are dominated by the semiclassical 
force~\cite{JoyceProkopecTurok:1995,dickwerk,ClineJoyceKainulainen:2000},
which is the leading order source for single 
fermions coupled to a space-time dependent background.  
 Unlike in some alternative approaches pursued in the literature, 
 a nice feature of the present approach is that sources and transport are 
treated within one formalism, which allows for
an unambiguous fixing of the amplitude of CP-violating sources 
in (diffusion) transport equations. Moreover, this approach
allows in principle for a systematic study of CP-violating sources from
collisions, and how thermal and off-shell effects 
may affect the sourcing and transport of CP-violating charges.

Furthermore, since our treatment is based on a formalism that fully includes
the effects of mixing fermions, our results are manifestly basis independent. 
This is in contrast to former work, where the transport was treated either 
in mass 
eigenbasis~\cite{JoyceProkopecTurok:1995,dickwerk,ClineJoyceKainulainen:2000},
or in flavour basis~\cite{Huet_Nelson,Madrid_group, Madrid_group_imp}, and
which describes just transport of two physical degrees of freedom,
ignoring any dynamical effects arising from flavour mixing.
For example, such a treatment of neutrino propagation
would lead to complete neglect of neutrino oscillations. 
Unlike in the neutrinos case, the chargino oscillations occur on a microscopic
scale given by the split in the chargino eigenvalues and by
the chargino damping. In addition our formalism contains genuinely 
sources and transport such that no phenomenological thermalization 
time $\tau$ has to be introduced as was done 
in~\cite{Huet_Nelson,Madrid_group, Madrid_group_imp}. 

 While a broad-brush picture of the first order sources resembles 
the sources found in Refs.~\cite{Madrid_group, Madrid_group_imp}
(this approach to supersymmetric baryogenesis was initiated
by Huet and Nelson~\cite{Huet_Nelson}), there are noteworthy differences. 
Firstly, we found that chargino flavour oscillations
are of crucial importance for identification 
and dynamics of the CP-violating sources. 
The oscillations tend to suppress the calculated baryon asymmetry,
in particular in the limit of a moderate damping, a feature that was not 
observed in~\cite{Madrid_group, Madrid_group_imp}. 
Secondly, while broadly speaking the first order sources
share similar parametric dependences with the earlier work,
they do differ in some important aspects. 

Firstly, as can be seen in Fig.~\ref{plot_ma_2}, all of the contributions
to the BAU from our first order sources are of similar size, such that 
in the final BAU one sees the characteristics of all three sources.
In particular, the BAU peaks at
$|\mu_c|\simeq |M_2|+20$~GeV, and then dies out rather fast for
large values of $|\mu_c|$. On the other hand, 
the BAU obtained in~\cite{Madrid_group, Madrid_group_imp}
peaks at the chargino mass degeneracy, $|\mu_c|\simeq |M_2|$, 
it is about a factor 2 larger than in our calculation, 
and finally it does not diminish 
for large values of $|\mu_c|$ as fast as in our calculation.
Both discrepancies are due to the oscillations. Far from degeneracy 
(large mass splitting $\Lambda$) the fast oscillations will suppress the 
particle densities. Near the degeneracy (small mass splitting $\Lambda$)
CP-violation is suppressed since it is generated by the oscillations 
as shown in Eq. (\ref{calc_equ}) and this suppression cancels the 
resonance in the sources observed in~\cite{Madrid_group_imp}.

 Provided it is not too strong, the phenomenological damping term
that we introduce does not significantly affect
the maximum strength of the first order sources, unlike what was observed
in~\cite{Madrid_group,Madrid_group_imp,BalazsCarenaMenonMorrisseyWagner:2004}.
On the other hand, the second order sources 
are enhanced  in the limit of a small damping,
as shown in Fig.~\ref{plot_source}. For example, 
for a moderate damping, $\Gamma\simeq \alpha_wT$, the first order sources 
dominate in most of the parameter space. The second order source is
small, such that that it cannot alone be a viable source for baryogenesis, 
even when the CP violation in the chargino sector is maximal.
When damping is weak,  $\Gamma\simeq 0.25\alpha_wT$, 
the second order source dominates in a large section of parameter space.
For even smaller values of $\Gamma$ the semiclassical force source alone
represents a viable baryogenesis source, implying that our source
is somewhat larger than what was found in
Ref.~\cite{ClineJoyceKainulainen:2000}, which 
agrees quite well with the BAU found in~\cite{Steffen:SEWMProceedings},
based on a study of semiclassical force source obtained in 
the mass eigenbasis~\cite{dickwerk}.

 Perhaps the most severe constraints on the supersymmetric baryogenesis 
in near future are expected from electric dipole moment (EDM) measurements.
Already the current constraints on the EDM of the 
electron~\cite{ReganComminsSchmidtDeMille:2002}
place rather strict constraints on the CP-violating phases
in the chargino mass matrix, as can be seen from Fig.~4 in 
Ref.~\cite{ChangChangKeung:2002} or Fig.~6 
in~\cite{Pilaftsis:2002fe} that claims a little less restrictive 
bounds. For example, for 
$\mu_c = 200$~GeV, $M_{H+} = 170$~GeV 
and $\tan(\beta) = 6$ the CP-phase is restricted to be less than
about $1/12$ and $1/10$, respectively, implying that, when our numbers are taken at
the face value, the baryogenesis mechanism presented here is by 
about {\it factor 5-6 too weak}
to account for the observed BAU. Similar conclusion
is reached for other values of $|\mu_c|$ and $|M_2|$ since both the EDMs and the
produced baryogenesis decreases with decreasing chargino masses.
We would like to emphasize that most of the parameters are chosen 
in order to produce as much baryon asymmetry as possible, {\it e.g.} the 
values used for the wall velocity $v_w$ and the wall width $L_w$. 
The only relevant parameter we have not varied so far is $\tan \beta$.
Smaller values of $\tan \beta$ lead to less restrictive EDMs and at the same
time to more baryon asymmetry as shown in Fig.~\ref{plot_beta_1}.
\begin{figure}[htbp]
\begin{center}
\epsfig{file=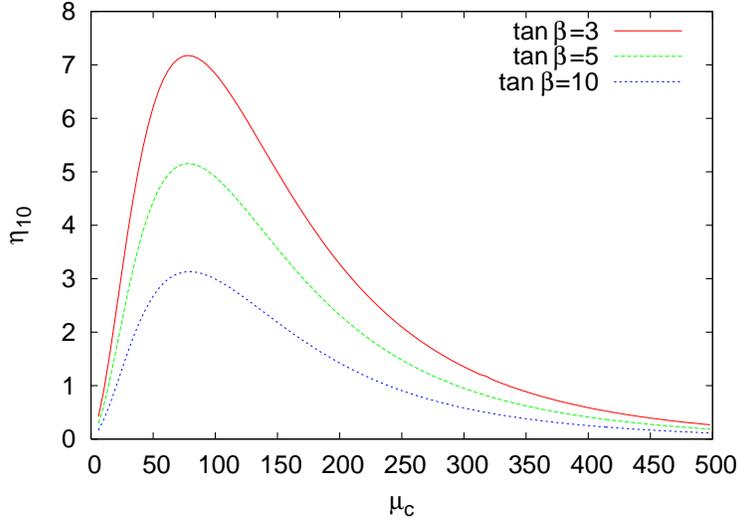, width=4.0 in}
\end{center}
\vskip -0.2in
\lbfig{plot_beta_1}
\caption[Fig:ma_1]{%
\small This plot shows $\eta_{10}=10^{10}\eta$ as a function of
$\mu_c$, $M_2=\mu_c - 20$ GeV, $m_A =150$ GeV and for several values of 
$\tan \beta$. }
\end{figure}
On the other hand $\tan \beta$ due to the mass of the lightest Higgs is restricted to
be in the range~\cite{Madrid_group}
\be
5 \lesssim \tan \beta 
\ee
such that we are not allowed to enter the region with smaller values of 
$\tan \beta$. In addition even for $\tan \beta=3$ our result is still a
factor 2 to small. 
Note that this is a very different conclusion from the one reached 
in Ref.~\cite{BalazsCarenaMenonMorrisseyWagner:2004},
where an ample region of parameter
space was claimed to result in a successful baryogenesis in the MSSM.

Based on our analysis, can we conclude that the MSSM baryogenesis 
is ruled out? At least a factor 2 can be accounted for based on 
the inaccuracies in the parameters in diffusion equations, 
as well as from approximations that lead to the set of 
diffusion equations considered here, but unlikely a 
factor 5~\cite{ClineJoyceKainulainen:2000}. 
Nevertheless, it would be premature to claim that the MSSM baryogenesis is 
ruled out, since the chargino mediated baryogenesis studied here does not 
exhaust the possibilities of the MSSM. Recall that neutralinos mediate
baryogenesis as well, and that their contribution may be as important
or even more important than that from charginos. 
Furthermore, in the complete set of diffusion equations, 
there may be additional channels, which lead to baryon production,
as of yet unaccounted for. 
Finally the EDM analysis given here is not 
conclusive. For larger values of $m_A$ and due to the possibility of
fortuitous cancellations between different EDM contributions, 
the value of the electron 
EDM could be reduced relative to the generic values used above
\footnote{C.E.M. Wagner, private communication.}.
Hence electroweak baryogenesis seems to be still possible in 
the MSSM and in this respect we agree with the conclusion 
drawn in~\cite{Madrid_group,Madrid_group_imp,BalazsCarenaMenonMorrisseyWagner:2004}.
However, we would like to emphasis two robust and novel 
consequences resulting from
the quantum treatment of transport in the chargino sector:
The BAU is strongly suppressed away from the chargino mass degeneracy
and one requires even close to the degeneracy a CP-violating phase of order 
unity, more precisely $\textrm{arg}(M_2 \mu_c) >0.2$.

Modifications of the MSSM with an additional singlet (NMSSM) also contain
the prominent chargino channels. As shown in~\cite{Huber:1998ck, Huber:1999sa}
 one easily can get a strong first order phase transition 
and also spontaneous CP-violation at the temperature $T_c$ of the 
phase transition not affecting the EDM at $T=0$.
This then allows for a satisfying baryon asymmetry without squeezing the 
(unfortunately many) parameters. One can also think about 
extensions of the MSSM that not forbid  
$\tan \beta \sim 1$~\cite{Han:2004yd} or modifications
of the Standard Model~\cite{Bodeker:2004ws}, where the chargino
system does not appear, but of course again quantum-transport is important.

 In summary, our numerical solution to 
the diffusion equations~(\ref{diff_Q}--\ref{diff_h}) 
shows that a successful baryogenesis at the electroweak scale from
charginos of the MSSM is possible only when CP violation is quite large,
and near the resonance, $|\mu_c - M_2| \ll 50 \textrm{ GeV}$, 
$M_2,\mu_c \ll 500$~GeV. 
As long as the first order sources dominate, due to the oscillations, 
this picture persists also for much stronger sources,
which is to be contrasted 
to~\cite{Madrid_group,Madrid_group_imp,BalazsCarenaMenonMorrisseyWagner:2004}.

Our conclusion is that, in purely chargino mediated MSSM baryogenesis
the capability to explain the observed baryon asymmetry is strongly constrained by the 
current electron EDM bounds. 

\section*{Acknowledgments}

We would like to thank M. Carena, M. Quiros and C. Wagner 
for useful comments. 
M.S. is supported by the Spanish Ministerio 
de Educaci\'on y Ciencia under grant EX2003-0696.

\appendix

\section{Comparison of Bosonic Sources\label{app}}

In this Appendix we show how the sources, presented in the
current work, relate with those of references 
\cite{Madrid_group,Madrid_group_imp} in the
bosonic case and in the limit of zero widths~\footnote{It has been
  checked that the numerical difference between the cases with and
  without finite width is smaller than $2-3\%$.}.  
In order to inspect this
we make use of the Kadanoff-Baym equations for the full
$2\times 2$ Green functions of the Schwinger-Keldysh formalism. 
These equations are obtained from Eqs.~(\ref{CE}--\ref{KE}) 
if we substitute $\Delta^<$ by the full $2\times 2$ propagator,
\begin{equation}
 \Delta = \left(\begin{array}{cc}
                            \Delta^{++} & \Delta^{+-} \cr           
                            \Delta^{-+} & \Delta^{--} \cr
                \end{array}
          \right)
        = \left(\begin{array}{cc}
                            \Delta^t & \Delta^< \cr           
                            \Delta^> & \Delta^{\bar t} \cr
                \end{array}
          \right)
\,,
\end{equation}
insert {\it unity} in the {\it r.h.s.}  of~(\ref{CE}), and
set the collision term of (\ref{KE}) to zero,
\begin{eqnarray}
\left(k^2-\frac{1}{4}\partial^2_X\right)\Delta
 - \frac12\left\lbrace {\cal M}^2,\Delta\right\rbrace
 + \frac{i}{4}\left\lbrack\partial_\mu^X{\cal M}^2,\partial_k^\mu \Delta
              \right\rbrack
  &=1
\label{A2}
\\
k\cdot\partial_X \Delta
 + \frac{i}{2}\left\lbrack {\cal M}^2,\Delta\right\rbrack
 +\frac{1}4\left\lbrace\partial_\mu^X{\cal M}^2,\partial_k^\mu \Delta
           \right\rbrace
 &=0
\,.
\label{A3}
\end{eqnarray}

 In \cite{Madrid_group} the corresponding Dyson-Schwinger equation
for $\Delta$ is iteratively solved as an expansion 
in powers of $\partial_\mu^X {\cal M}^2$,
\begin{equation}
 \Delta = \Delta^{(0)} + \Delta^{(1)} + \dots 
\label{Delta-0-1}
\end{equation}
where $\Delta^{(0)}$ is the leading order equilibrium propagator,
and $\Delta^{(1)} = {\cal O}(\partial_\mu^X {\cal M}^2)$
denotes a first order correction.
 Upon performing a Wigner transform over the spatial variables,
$\int d^4(x-y)\,{\rm e}^{i(x-y)\cdot k}$, identifying $z=(x+y)/2\equiv X$, 
and transforming into the flavour basis, 
the first order correction, $\Delta^{(1)}=\Delta^{(1)}(k;X)$ given in
Eq.~(2.5) of Ref.~\cite{Madrid_group} becomes
\begin{equation}
\Delta^{(1)}=\frac{i}{2}\left[(\partial_k^\mu\Delta^{(0)})
(\partial_\mu^X{\cal M}^2)\Delta^{(0)}-\Delta^{(0)}
(\partial_\mu^X{\cal M}^2)\partial_k^\mu\Delta^{(0)}\right]
\,.
\label{Delta-1}
\end{equation}

In the approach advocated in~\cite{Madrid_group} 
in the calculation of the sources one is not interested 
in long range effects, and hence the term  $\partial_X^2 \Delta^{(1)}$ 
in the constraint equation~(\ref{A2}) and $k\cdot \partial_X \Delta^{(1)}$ 
in~(\ref{A3}) were considered of second order, and thus have been neglected.
 From Ref.~\cite{fermions} and this work we know however 
that, when the dynamics is taken account of, 
in the case of mixing scalars and fermions the flavour oscillations mess up 
the derivative expansion, such that only the terms 
containing spatial derivatives acting on the mass term are genuinely 
derivative-suppressed.

 Upon inserting~(\ref{Delta-0-1}) into~(\ref{A2}--\ref{A3}) 
and using the prescription for derivative counting of~\cite{Madrid_group},
we get for the leading order propagator,
\begin{equation}
 k^2\Delta^{(0)} - \frac 12 \big\{{\cal M}^2,\Delta^{(0)}\big\} = 1 
\,,
\label{Delta-0}
\end{equation}
which is solved by the thermal Green function,
which commutes with ${\cal M}^2$. The first order equations are,
\begin{eqnarray}
   k^2\Delta^{(1)}
 - \frac12\left\lbrace {\cal M}^2,\Delta^{(1)}\right\rbrace
 + \frac{i}{4}\left\lbrack \partial_\mu^X{\cal M}^2,\partial_k^\mu\Delta^{(0)}
           \right\rbrack
   &=& 0
\label{cons1}
\\
   k\cdot\partial_X \Delta^{(0)}
  + \frac{i}{2}\left\lbrack{\cal M}^2,\Delta^{(1)}\right\rbrack
  + \frac{1}4\left\lbrace\partial_\mu^X{\cal M}^2,\partial_k^\mu \Delta^{(0)}
             \right\rbrace
 &=&0
\label{kin1}
\,.
\end{eqnarray}
By a judicious use of~(\ref{Delta-0}) and its derivatives, 
\begin{eqnarray}
 (k^2 - {\cal M}^2)\Delta^{(0)} &=& 1 = \Delta^{(0)} (k^2 - {\cal M}^2)
\nonumber\\
 (k^2 - {\cal M}^2)\partial_\mu^X\Delta^{(0)}
              &=& (\partial_\mu^X{\cal M}^2)\Delta^{(0)} 
\,,\qquad 
(\partial_\mu^X\Delta^{(0)}) (k^2 - {\cal M}^2)
              = \Delta^{(0)} (\partial_\mu^X{\cal M}^2)
\nonumber\\
 (k^2 - {\cal M}^2)\partial^\mu_k\Delta^{(0)}
              &=& -(2k^\mu)\Delta^{(0)} 
               = (\partial^\mu_k\Delta^{(0)}) (k^2 - {\cal M}^2)
\,
\end{eqnarray}
one finds that the first order correction~(\ref{Delta-1}) can be recast as, 
\begin{eqnarray}
\Delta^{(1)}&=& \frac{i}{2}\left[(\partial_k^\mu\Delta^{(0)})
               (k^2-{\cal M}^2)\partial_\mu^X\Delta^{(0)}
               -(\partial_\mu^X\Delta^{(0)})(k^2-{\cal M}^2)
                    \partial_k^\mu\Delta^{(0)}\right]
\nonumber\\
            &=& -i\left[\Delta^{(0)} k\cdot\partial_X\Delta^{(0)}
                       -(k\cdot\partial_X\Delta^{(0)})\Delta^{(0)}\right]
\,.
\label{Delta-1:2}
\end{eqnarray}
It can be easily shown that, when this is inserted 
into~(\ref{cons1}--\ref{kin1}), 
one obtains two consistent equations for $\Delta^{(1)}$. 
%However $\Delta^{(1)}$ is not a solution of our transport and 
%constraint equations, 
%even in the limit
%$\Gamma_h \to 0$, since the terms $k\cdot \partial_X \Delta^{(1)}$ and 
%$\partial_X^2 \Delta^{(1)}$ have been neglected.

 Note that taking moments of the kinetic
equation~(\ref{kin1}) allows for a simple prescription 
on how the CP-violating source originally calculated in~\cite{Madrid_group}
enters the relevant transport equations for squarks. 
The term $\Delta^{(1)}$
enters through the commutator $[{\cal M}^2,\Delta^{(1)}]$ in~(\ref{kin1}),
while~\cite{Madrid_group} used a heuristic prescription for the sources 
based on the Fick's law and interpreted the diagonal entries of $\Delta^{(1)}$
in the interaction basis as sources for the classical diffusion equations. 

 Note further that, even though we have rephrased the source 
of~\cite{Madrid_group} in our language, it remains a nontrivial matter
to establish the exact correspondence between the source
of~\cite{Madrid_group} appearing in~(\ref{kin1}) 
and the source calculated in this work.
Our source is in principle obtained by the means of the kernel of
Eq.~(\ref{calc_equ}) acting upon~(\ref{bos_src}),
which is thus of a complicated nonlocal form, and bares no simple relation to 
the source in~(\ref{kin1}), apart from a rather superficial similarity,
stemming from the fact that the kernel of
Eq.~(\ref{calc_equ}) is a nonlocal functional of the commutator
$[{\cal M}^2(z^\prime),\cdot ]$ acting upon~(\ref{bos_src})
(see Ref.~\cite{fermions}).

Finally, we emphasize that the difference in how 
the sources couple to the diffusion equations cannot alone 
explain a different baryon asymmetry obtained by the two methods,
but also the presence of the oscillatory terms. 

As regards the case of mixing fermions, we expect that the sources
can be related in a similar fashion. Because of the spinor structure however,
the comparison for fermions is a nontrivial generalization of the bosonic case.

\end{document}